\pdfoutput=1
\documentclass[aps,prl,reprint]{revtex4-1}

\usepackage{amssymb,amsmath}
\usepackage{graphicx}

\newcommand{\opn}[1]{\operatorname{#1}}

\newcommand{\wtil}[1]{\widetilde{#1}}

\newcommand{\imag}{\opn{Im}}

\newcommand{\der}[2]{\frac{\partial #1}{\partial #2}}

\newcommand{\jt}{\textstyle}

\newcommand{\IGNORE}[1]{}

  \def\XXint#1#2#3{{\setbox0=\hbox{$#1{#2#3}{\int}$}
      \vcenter{\hbox{$#2#3$}}\kern-.5\wd0}}

\begin{document}

\title{Breakdown of Self-Similarity at the Crests of Large-Amplitude Standing Water Waves}

\author{Jon Wilkening} \email{wilken@math.berkeley.edu}
\affiliation{Department of Mathematics, University of California, Berkeley, California, 94720, USA} 

\begin{abstract} 
  We study the limiting behavior of large-amplitude standing waves on
  deep water using high-resolution numerical simulations in double and
  quadruple precision. While periodic traveling waves 
  approach Stokes's sharply crested extreme wave in an asymptotically
  self-similar manner, we find that standing waves behave differently.
  Instead of sharpening to a corner or cusp as previously conjectured,
  the crest tip develops a variety of oscillatory structures.  This
  causes the bifurcation curve that parametrizes these waves to fragment
  into disjoint branches corresponding to the different oscillation
  patterns that occur.  In many cases, a vertical jet of fluid pushes
  these structures upward, leading to wave profiles commonly seen in
  wave tank experiments.  Thus, we observe a rich array of dynamic
  behavior at small length scales in a regime previously thought to be
  self-similar.
\end{abstract}

\maketitle


Singularities in fluid mechanics are generally expected to be
asymptotically self-similar \cite{eggers:09}.  These can be dynamic
singularities, such as bubble pinch-off \cite{turitsyn:09} or wave
breaking \cite{bridges:09}, or parametric singularities, where a
family of smooth solutions terminates at a singular solution.  A
famous example of the latter type was posed by Stokes in 1880, who
used an asymptotic expansion of the stream function to argue that the
periodic traveling water wave of greatest height should have an
interior crest angle of $120^\circ$.  This crest angle has been
confirmed in numerous computational studies \cite{gandzha:07} as well
as theoretically \cite{amick:82}.  The asymptotic behavior of the
almost highest traveling wave was analyzed by Longuet-Higgins and Fox
\cite{lhf:77,lhf:78}.


Because genuine dynamics are involved, existing numerical methods have
been unable to maintain the accuracy needed to fully explore the
limiting behavior of large-amplitude standing waves.  As a result,
Penney and Price's conjecture \cite{penney:52} that a limiting
standing wave exists and develops $90^\circ$ interior crest angles
each time the fluid comes to rest has remained open since 1952.  Such
a singularity would be both dynamic and parametric.  The standing
waves in question are spatially periodic and have zero impulse
(horizontal momentum), maintaining even symmetry for all time.  They
are also temporally periodic, alternately passing through two
zero-velocity rest states of maximal potential energy.

\begin{figure}[b]
\includegraphics[width=3.4in]{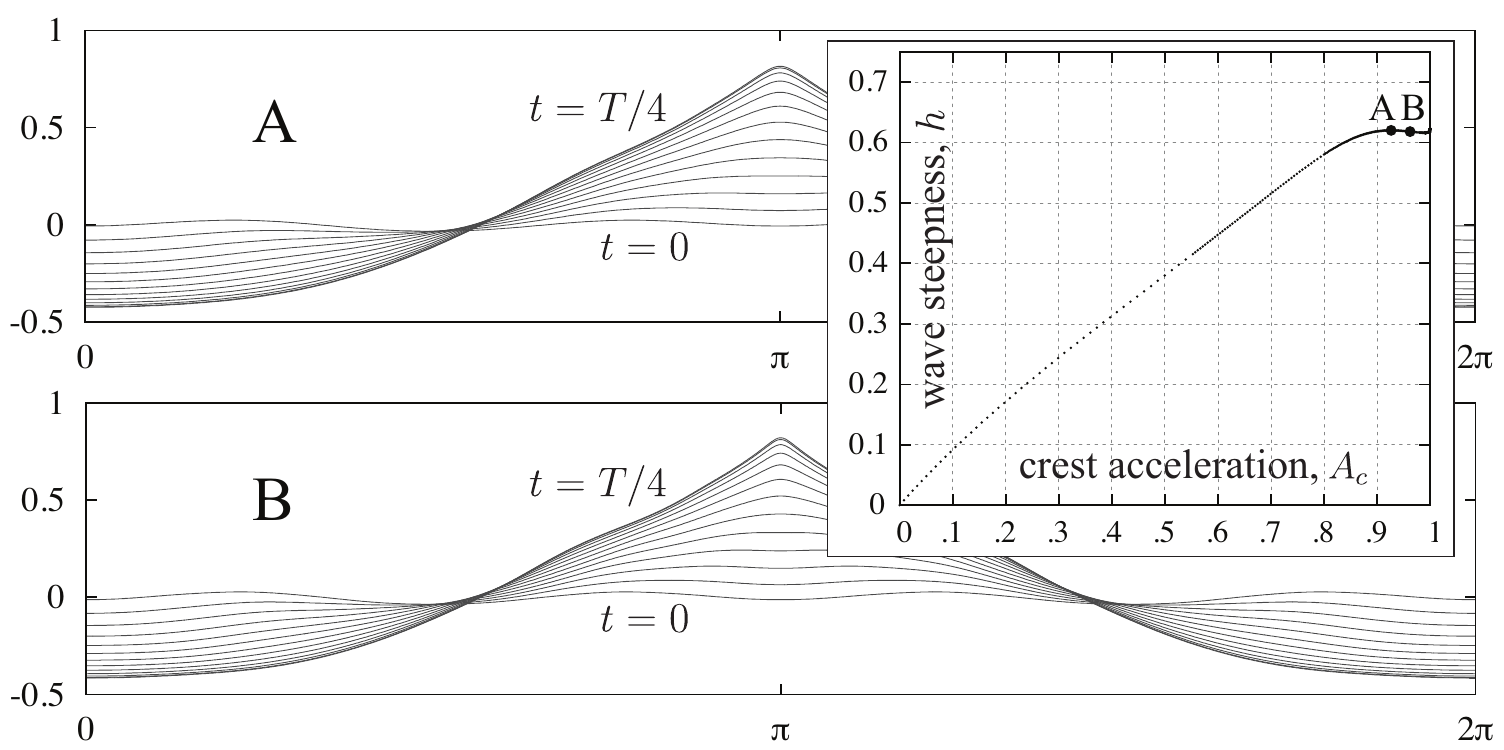}
\caption{\label{fig:mercer} Bifurcation diagram and selected standing
  waves, plotted at equal time slices over a quarter period.  The
  wavelength is taken to be $2\pi$, and $g=1$.  The crest tip sharpens
  as $A_c$ increases over the range $0\le A_c\le 0.985$, where
  previous numerical studies are reliable. In particular, the
  curvature at the crest is visibly higher for solution B than for A.
}
\end{figure}

Small-amplitude standing waves of this type were proved to exist by
Iooss, Plotnikov, and Toland \cite{iooss05}.  Larger-amplitude waves
were computed by Mercer and Roberts \cite{mercer:92}, who discovered
that the wave steepness (half the crest-to-trough height) does not
increase monotonically over the entire one-parameter family of
standing waves.  They proposed using (downward) crest acceleration,
$A_c$, as a continuation parameter instead.  We reproduce (and extend)
their plot of wave steepness versus crest acceleration in
Fig.~\ref{fig:mercer}.  Since pressure increases with depth near the
free surface \cite{wu:97}, Euler's equations imply that $A_c$ cannot
exceed $g$, the acceleration of gravity.

Taylor \cite{taylor:53} performed wave tank experiments and confirmed
that large-amplitude standing waves do form reasonably sharp crests
close to 90 degrees. A further increase in amplitude caused the waves
to splash and become unstable in the transverse direction.  Grant
\cite{grant} and Okamura \cite{okamura:98} have written theoretical
papers to support the $90^\circ$ conjecture. Okamura also performed
numerical experiments \cite{okamura:03,okamura:10} to back this claim.
Extrapolating from numerical solutions, Mercer and Roberts
\cite{mercer:92} speculated that the limiting crest angle might be as
sharp as $60^\circ$.  Schultz et.~al.~\cite{schultz} also predicted a
limiting wave profile with a crest angle smaller than $90^\circ$ and
offered the possibility that a cusp may form instead of a corner.

Our objective is to challenge the assumption that standing waves
behave as traveling waves in their approach of an ``extreme'' limiting
wave.  If there is no limiting wave profile, then a local analysis
suggesting a geometric singularity (corner or cusp) is inapplicable.

The equations of motion for a two-dimensional irrotational ideal fluid
of infinite depth are
\begin{subequations}
\label{eq:water}
\begin{align}
\label{eq:water:e}
  \eta_t &= \phi_y - \eta_x\phi_x, \\
\label{eq:water:f}
  \Phi_t &= P\left[ \phi_y\eta_t - \jt\frac{1}{2}
    |\nabla\phi|^2 - g\eta \right],
\end{align}
\end{subequations}
where $\eta(x,t)$ is the upper boundary of the evolving fluid and
$\Phi(x,t)=\phi(x,\eta(x,t),t)$ is the restriction of the velocity
potential to the free surface.  Both $\eta(x,t)$ and $\Phi(x,t)$ are
assumed to be $2\pi$ periodic in~$x$. In (\ref{eq:water:f}), $P$ is
the orthogonal projection to zero mean.  This equation comes from
$\Phi_t = \phi_t + \phi_y\eta_t$ and the unsteady Bernoulli equation
$\phi_t + \frac{1}{2}|\nabla\phi|^2 + \frac{p}{\rho} + gy = c(t)$,
where the arbitrary constant $c(t)$ is chosen to preserve the mean of
$\Phi(x,t)$.

To evaluate the right-hand side of (\ref{eq:water}) for the purpose of
time stepping, we use a boundary integral collocation method.  Details
will be given elsewhere \cite{jia:11}.  Briefly, we represent $\phi$
at a point $z=x+iy$ in the fluid using a double layer potential.
Suppressing $t$ in the notation and summing over periodic images
\cite{vtxs1}, the result is
\begin{equation}\label{eq:phi:layer}
  \phi(z) = \frac{1}{2\pi}
  \int_0^{2\pi} \wtil K(z,\alpha)\mu(\alpha)\,d\alpha,
\end{equation}
where $\wtil K(z,\alpha) = \imag\big\{\frac{\zeta'(\alpha)}{2}
  \cot\big(\frac{z - \zeta(\alpha)}{2}\big)\big\}$.  A
prime represents a derivative with respect to $\alpha$, and
\begin{equation}\label{eq:zeta}
  \zeta(\alpha) = \xi(\alpha) + i\eta(\xi(\alpha))
\end{equation}
is a parametrization of the curve.  The change of variables
$x=\xi(\alpha)$ allows for smooth mesh refinement near the crest tip.
Letting $z$ approach the boundary, we obtain a second-kind Fredholm
integral equation for $\mu$:
\begin{gather}
\label{eq:fred}
  \Phi(\xi(\alpha)) =
  \frac{\mu(\alpha)}{2} + \frac{1}{2\pi}
  \int_0^{2\pi}K(\alpha,\beta)\mu(\beta)\,d\beta, \\
\notag
  K = \imag\left\{
    \frac{\zeta'(\beta)}{2}
    \cot\left(\frac{\zeta(\alpha)-\zeta(\beta)}{2}\right) -
    \frac{1}{2}\cot\left(\frac{\alpha - \beta}{2}\right)
  \right\}.
\end{gather}
Once $\mu(\alpha)$ is known, we compute
$\phi_x$ and $\phi_y$ on the boundary from (\ref{eq:phi:layer}),
closing the system (\ref{eq:water}); see \cite{jia:11}.

We discretize space and time adaptively to resolve the solution as it
becomes increasingly singular.  Time is divided into $\nu$ segments
$\theta_lT$, where $\theta_1 + \cdots + \theta_\nu = 1/4$ and $T$ is
the current guess for the period.  On segment $l$, we fix the number
of (uniform) time steps, $N_l$, the number of spatial grid points,
$M_l$, and the function
\begin{equation*}
  \xi_l(\alpha) = \int_0^\alpha E_l(\beta)\,d\beta, \quad
  E_l(\alpha) = 1 - P\left[A_l\sin^4(\alpha/2)\right],
\end{equation*}
which controls the grid spacing in the change of variables
$x=\xi_l(\alpha)$.  $A_l$ is a parameter chosen between 0 (uniform
spacing) and 8/5, the value where $\xi_l(\alpha)$ ceases to be a
diffeomorphism.  As before, $P$ projects out the mean.

\begin{figure}[b]
\includegraphics[width=3.2in,trim=0 3 0 2,clip]{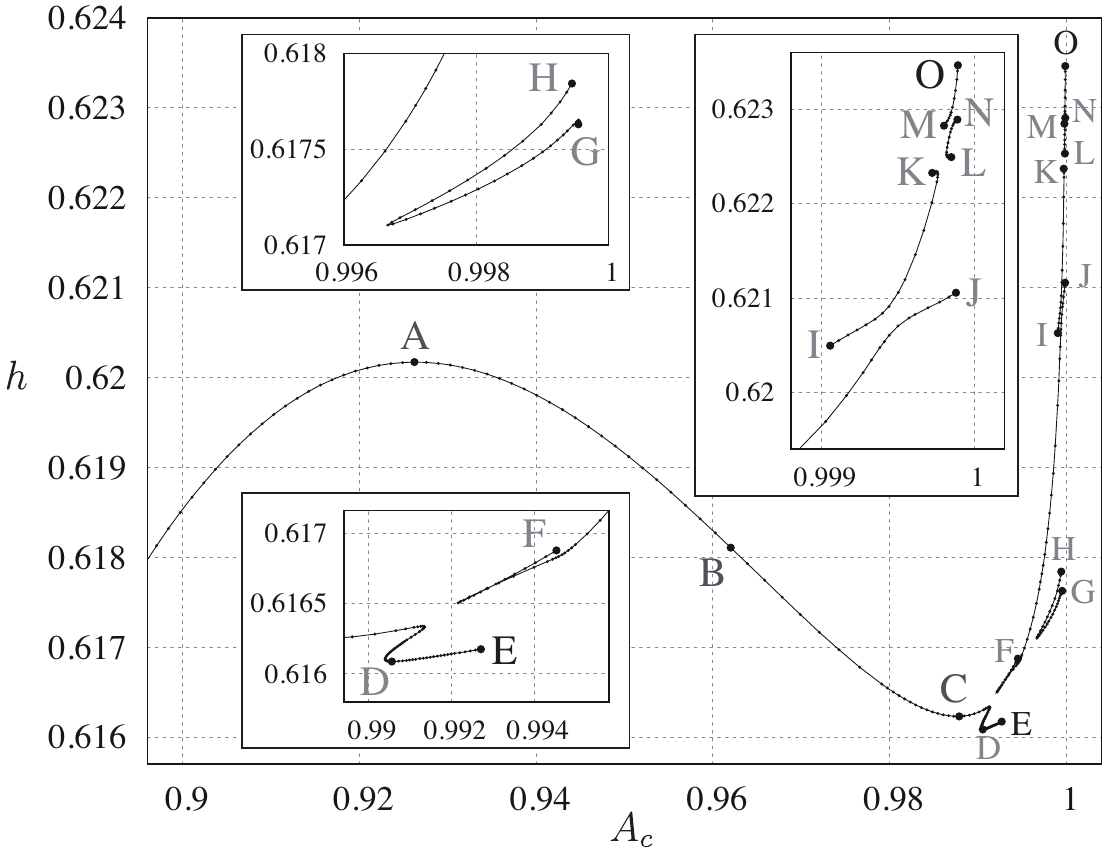}
\caption{\label{fig:bif2} The bifurcation curve in
  Fig.~\ref{fig:mercer} becomes fragmented in the range
  $0.985<A_c<1$, where previous numerical studies break down.  The
  labels A--O correspond to wave profiles shown in
  Fig.~\ref{fig:pan:zoom}.  The turning point in wave steepness at
  C, the lack of monotonicity in $A_c$, the complicated branching
  structure, and the existence of standing waves with $h>0.62017$ were
  not previously known.  }
\end{figure}

To compute standing waves, we use the Levenberg-Marquardt method
\cite{nocedal}, a trust-region algorithm for nonlinear least squares
problems, to minimize
\begin{equation}\label{eq:f}
  f(c) = \frac{1}{4\pi}\int_0^{2\pi}\Phi(x,T/4)^2\,dx, \qquad
  c\in R^{n+1},
\end{equation}
where $c$ contains the period as well as the nonzero Fourier
modes of the initial conditions; i.e.,~$T=c_0$ and
\begin{equation}\label{eq:ic}
  \hat\eta_k(0) = c_{|k|} \quad (k\text{ odd}), \quad\;
  \hat\Phi_k(0) = c_{|k|} \quad (k\text{ even}).
\end{equation}
Here $k$ ranges from $-n$ to $n$, excluding 0, and $n$ is chosen to be
close to $\frac{1}{4}M_1$, leaving the upper half of the spectrum of
$\eta$ and $\Phi$ to be zero initially.  A symmetry argument
\cite{mercer:92} shows that driving the velocity potential to zero at
time $T/4$ with initial conditions of the form (\ref{eq:ic}) leads to
a standing wave with period $T$ and zero impulse.  The method fails if
$f$ reaches a nonzero local minimum.

We discretize (\ref{eq:f}) with spectral accuracy by redefining
$f=\frac{1}{2}r^Tr$, where $r\in\mathbb{R}^m$, $m=M_\nu$, and
\begin{equation*}
  r_i = \Phi(\xi_\nu(\alpha_i),T/4)\sqrt{E_\nu(\alpha_i)/m}, \qquad
  \alpha_i = 2\pi i/M_\nu.
\end{equation*}
The square root comes from $dx=E_\nu(\alpha)\,d\alpha$.  Typically,
$4n\le m\le10n$.  To track families of solutions, one of the $c_k$ is
chosen as a continuation parameter \cite{keller:87} and eliminated
from the search space when minimizing $f$.  When a turning point is
detected in this $c_k$, we switch to a different one; see
\cite{vtxs1,jia:11} for details. The Jacobian $J_{ik}= \partial
r_i/\partial c_k$ is computed by solving the linearization of
(\ref{eq:water}) about the current solution to obtain
$\der{}{c_k}\Phi(x,T/4)$.  This can be parallelized very efficiently
\cite{jia:11}, dramatically increasing the resolution we are able to
achieve.

\begin{figure*}
  \includegraphics[width=6.26in]{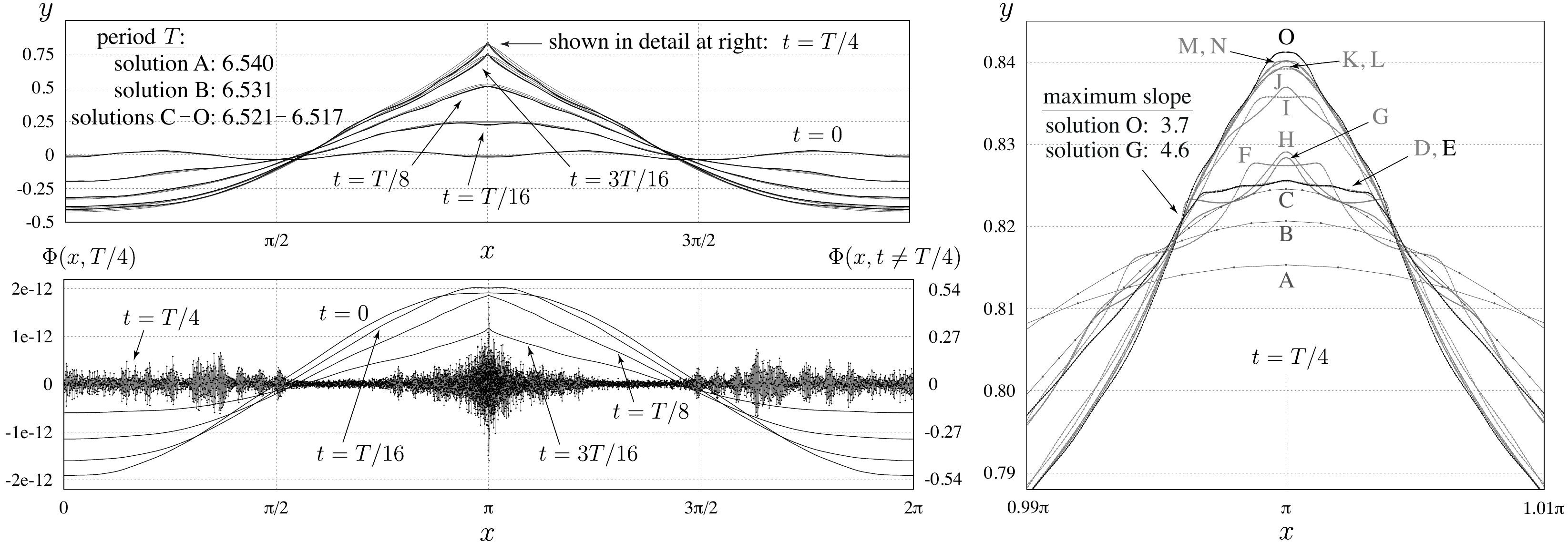}
  \caption{\label{fig:pan:zoom} Evolution of standing waves and
    velocity potential over a quarter period.  (Top left) Solutions
    A--O in Fig.~\ref{fig:bif2} are plotted on top of each other at
    the indicated times.  (Right) These solutions develop oscillatory
    structures near the crest that change phase across disconnections
    in the bifurcation diagram.  Solutions D--O have 350--600 grid
    points between $0.99\pi$ and $1.01\pi$. With at most 3 grid points
    in this interval, previous numerical studies could not resolve
    these structures.  (Bottom left) The velocity potential of
    solution O has been driven almost to zero at $t=T/4$, yielding
    $f=1.3\times 10^{-26}$.  For this solution, we used $\nu=4$,
    $\theta_l=\{0.2,0.2,0.4,0.2\}$, $M_l=\{4608,6144,6912,8192\}$,
    $N_l=\{192,288,768,480\}$, and $A_l=\{0,0.774,1.358,1.381\}$.  }
\end{figure*}


Our results are summarized in Figs.~\ref{fig:bif2}
and~\ref{fig:pan:zoom}.  First, we corroborate the result of Mercer
and Roberts \cite{mercer:92} that wave steepness, $h$, reaches a local
maximum of $h_\text{max}=0.62017$ at $A_c=0.92631$.  (The values
reported in \cite{mercer:92} were $0.6202$ and $0.9264$.)  Using
quadruple precision, we are able to compute $h_\text{max}$ to 26
digits of accuracy and the corresponding $A_c$ to 13 digits.  Okamura
\cite{okamura:03}, who found that $h$ increases monotonically all the
way to $A_c=1$, was incorrect.  Second, we find that crest
acceleration has turning points at $A_c=0.99135$ and $0.99040$.  This
is a surprise, as $A_c$ was chosen as a continuation parameter in
\cite{mercer:92} to avoid the lack of monotonicity in $h$. In our
work, $h$ and $A_c$ are plotted parametrically as functions of
whichever $c_k$ is currently used as a continuation parameter. Finally,
in the process of tracking this primary branch of solutions, we
discovered several other families of standing waves.  Each of these
branches was tracked in both directions until the computations became
too expensive to continue further with the desired accuracy,
$f\sim10^{-26}$ in double precision.

\begin{figure}[b]
\includegraphics[width=3.4in,trim=0 2 0 1,clip]{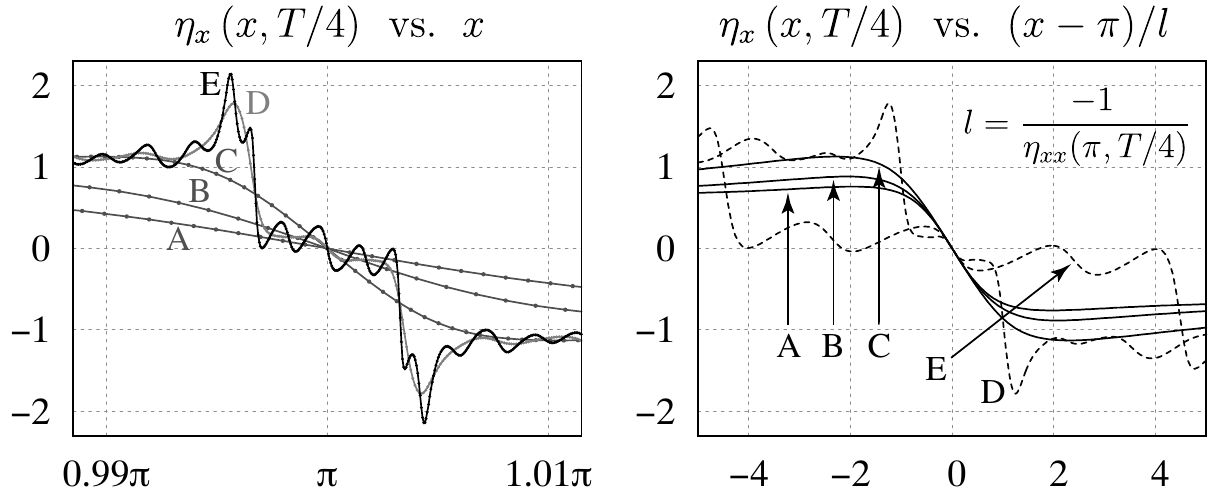}
\caption{ \label{fig:slopes} Breakdown of self-similarity on the
  primary branch.  When lengths are rescaled so the radius of
  curvature at the crest is~1, the slopes of solutions A--C have
  similar shapes.  In the traveling case (Fig.~\ref{fig:trav}), these
  rescaled slopes would approach a limiting curve.  But, for standing
  waves, oscillations develop, and a limiting curve does not emerge.  }
\end{figure}

The standing waves that constitute these branches look qualitatively
similar to each other in the large, where they closely resemble the
photographs from Taylor's wave tank experiments \cite{taylor:53}.
However, as illustrated in Fig.~\ref{fig:pan:zoom}, solutions on
different branches feature different oscillation patterns in the
vicinity of the crest tip.  The rapid increase in wave steepness from
solution E to solution O in Fig.~\ref{fig:bif2} corresponds to a
vertical jet of fluid that forms near the crest before the standing
wave reaches its rest state.  The resulting protrusion causes the
maximum slope to be much larger than 1 for most of these solutions.
Taylor photographed similar structures at the crest in his wave tank
experiments.  Schultz et.~al.~\cite{schultz} argued that surface
tension was responsible for these protrusions, but we find that they
occur even without surface tension.  Comparing solutions A--E on the
primary branch, we see that solutions eventually flatten out at the
crest and become oscillatory rather than sharp.  Figure~\ref{fig:slopes}
provides further evidence that these oscillations grow large enough to
prevent this family of solutions from approaching a limiting wave
profile in an asymptotically self-similar fashion.

\begin{figure*}
\includegraphics[width=6.8in,trim=0 3 0 2,clip]{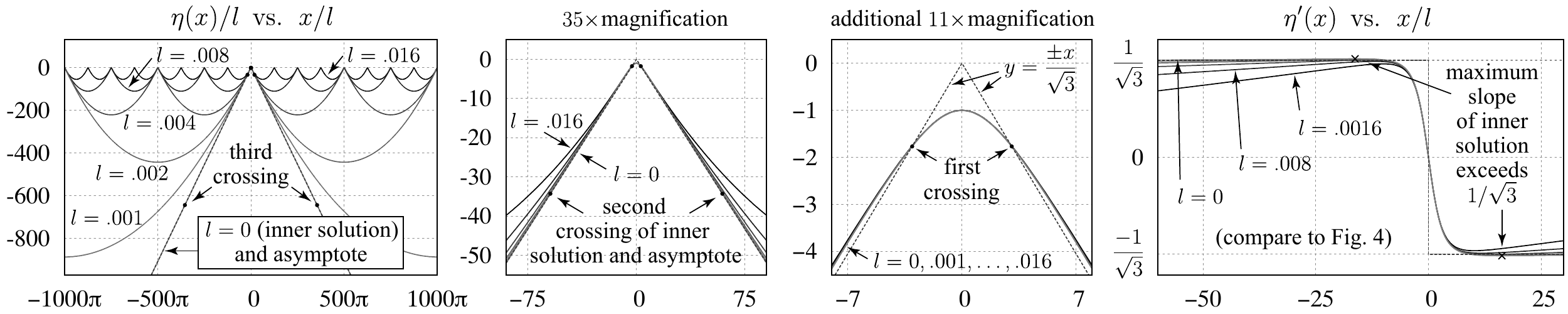}
\caption{\label{fig:trav} \hspace*{-5pt}
  Self-similar asymptotics of traveling water
  waves near the wave crest. (Left) Using a variant of our standing
  wave code, we computed 5 periodic traveling solutions with
  wavelength $2\pi$ and particle speed $q$ at the wave crest, chosen
  so that $l=q^2/2g$ has the values shown.  We then plotted
  $\eta(x)/l$ versus $x/l$, as well as the inner solution of
  \cite{lhf:77}.  The distance between successive crossings of the
  inner solution with its asymptote grows exponentially; thus, $l$
  must be extremely small to observe oscillatory behavior near the
  crest.  (Right) As $l\rightarrow0$, the slopes of the rescaled
  periodic waves approach the slope of the inner solution.  }
\end{figure*}

Regarding accuracy, our method is spectrally accurate in space, 8th or
15th order in time \cite{jia:11}, and quadratically convergent in the
search for a minimizer of $f$ in (\ref{eq:f}).  We achieve robustness
by formulating the shooting method as an overdetermined nonlinear
least squares problem.  If the numerical solution loses resolution,
the equations $r_i(c)=0$ become incompatible with each other and the
objective function $f=\frac{1}{2}r^Tr$ grows accordingly.  This
prevents the method from giving misleading overestimates of the
accuracy of the standing waves it finds.  For example, we recomputed
solution O of Fig.~\ref{fig:pan:zoom} in quadruple precision on a
finer mesh ($M_l=\{6144,7500,8192,9216\}$, $\theta_l=\{0.1,0.3,0.4,0.2\}$,
and $A_l=\{0,1.043,1.405,1.476\}$), using the initial conditions
obtained by minimizing $f$ in double precision.  The more accurately
computed value of $f$ is $8.6\times10^{-27}$, which is 34\% smaller
than predicted in double precision.  This level of inaccuracy in the
predicted error is acceptable, as driving $\Phi(x,T/4)$ to zero entails
eliminating as many significant digits as possible. For solution A, we
repeated the minimization in quadruple precision, causing $f$ to
decrease from $2.2\times 10^{-28}$ to $2.1\times 10^{-60}$.  In
addition to $f$, we monitor energy conservation and the decay of
Fourier modes at various times to ensure that $\eta$ and $\Phi$ remain
resolved to machine precision; see~\cite{jia:11} for more details.

It is instructive to compare our results to the traveling wave case.
Longuet-Higgins and Fox \cite{lhf:77,lhf:78} showed that periodic traveling
waves are asymptotically self-similar in two scaling regimes.  If the
wavelength, $L$, is held fixed as the crest tip sharpens, the limiting
wave profile has a $120^\circ$ corner.  This is the outer solution of
\cite{lhf:78}, predicted by Stokes and proved to exist in
\cite{amick:82}.  If, instead, the fluid velocity at the crest remains
fixed as the wavelength goes to infinity, the limiting wave profile is
shown in Fig.~\ref{fig:trav}.  This inner solution crosses the
asymptotes $y=\pm x/\sqrt{3}$ infinitely often \cite{lhf:77}, implying
that traveling waves approach Stokes's limiting wave in an oscillatory
manner, rather than monotonically, with $L$ fixed.

The oscillations in the standing wave case are of a completely
different nature.  No choice of scaling will cause the curves in
Fig.~\ref{fig:slopes} to approach a limiting inner solution.  We
believe these oscillations are caused by resonant modes in the
two-point boundary value problem (\ref{eq:water}) with boundary
conditions $\Phi(x,\pm T/4)=0$, treating $T$ as a bifurcation
parameter.  A resonant mode is a perturbation that nearly satisfies
the linearized boundary value problem.  Such modes can be strongly
excited in the process of computing standing waves, especially in
finite depth \cite{mercer:94,smith:roberts:99,jia:11}.  Disconnections
in the bifurcation diagram seem to occur when a resonant mode can be
excited with more than one amplitude. For example, solutions I and J
in Fig.~\ref{fig:pan:zoom} both contain a secondary, higher-frequency
standing wave (the resonant mode) superimposed on a low-frequency
carrier wave. The secondary wave sharpens the crest at J and flattens
it at I, being 180 degrees out of phase from one branch to the other.

We conclude that resonance is responsible for oscillations and trumps
self-similarity in determining the dynamics of standing waves at small
scales.  This shows that, although under-resolved numerical simulations
may exhibit self-similar dynamics, as happened in \cite{okamura:03},
the true dynamics may be more complex.  Recent work on singularity
formation in free surface flow problems, such as droplet and bubble
pinch-off \cite{eggers:09, turitsyn:09} and wave breaking
\cite{bridges:09}, may also benefit from higher-resolution
simulations, which could reveal new aspects of their dynamics.

\begin{acknowledgments}
  This research was supported by the National Science Foundation
  (DMS-0955078) and the U.S. Department of Energy (DE-AC02-05CH11231).
  The computations were performed on the Lawrencium cluster at LBNL.
\end{acknowledgments}


\begin{thebibliography}{23}%
\makeatletter
\providecommand \@ifxundefined [1]{%
 \@ifx{#1\undefined}
}%
\providecommand \@ifnum [1]{%
 \ifnum #1\expandafter \@firstoftwo
 \else \expandafter \@secondoftwo
 \fi
}%
\providecommand \@ifx [1]{%
 \ifx #1\expandafter \@firstoftwo
 \else \expandafter \@secondoftwo
 \fi
}%
\providecommand \natexlab [1]{#1}%
\providecommand \enquote  [1]{``#1''}%
\providecommand \bibnamefont  [1]{#1}%
\providecommand \bibfnamefont [1]{#1}%
\providecommand \citenamefont [1]{#1}%
\providecommand \href@noop [0]{\@secondoftwo}%
\providecommand \href [0]{\begingroup \@sanitize@url \@href}%
\providecommand \@href[1]{\@@startlink{#1}\@@href}%
\providecommand \@@href[1]{\endgroup#1\@@endlink}%
\providecommand \@sanitize@url [0]{\catcode `\\12\catcode `\$12\catcode
  `\&12\catcode `\#12\catcode `\^12\catcode `\_12\catcode `\%12\relax}%
\providecommand \@@startlink[1]{}%
\providecommand \@@endlink[0]{}%
\providecommand \url  [0]{\begingroup\@sanitize@url \@url }%
\providecommand \@url [1]{\endgroup\@href {#1}{\urlprefix }}%
\providecommand \urlprefix  [0]{URL }%
\providecommand \Eprint [0]{\href }%
\providecommand \doibase [0]{http://dx.doi.org/}%
\providecommand \selectlanguage [0]{\@gobble}%
\providecommand \bibinfo  [0]{\@secondoftwo}%
\providecommand \bibfield  [0]{\@secondoftwo}%
\providecommand \translation [1]{[#1]}%
\providecommand \BibitemOpen [0]{}%
\providecommand \bibitemStop [0]{}%
\providecommand \bibitemNoStop [0]{.\EOS\space}%
\providecommand \EOS [0]{\spacefactor3000\relax}%
\providecommand \BibitemShut  [1]{\csname bibitem#1\endcsname}%
\let\auto@bib@innerbib\@empty
\bibitem [{\citenamefont {Eggers}\ and\ \citenamefont
  {Fontelos}(2009)}]{eggers:09}%
  \BibitemOpen
  \bibfield  {author} {\bibinfo {author} {\bibfnamefont {J.}~\bibnamefont
  {Eggers}}\ and\ \bibinfo {author} {\bibfnamefont {M.~A.}\ \bibnamefont
  {Fontelos}},\ }\href@noop {} {\bibfield  {journal} {\bibinfo  {journal}
  {Nonlinearity}\ }\textbf {\bibinfo {volume} {22}},\ \bibinfo {pages} {R1}
  (\bibinfo {year} {2009})}\BibitemShut {NoStop}%
\bibitem [{\citenamefont {Turitsyn}\ \emph {et~al.}(2009)\citenamefont
  {Turitsyn}, \citenamefont {Lai},\ and\ \citenamefont {Zhang}}]{turitsyn:09}%
  \BibitemOpen
  \bibfield  {author} {\bibinfo {author} {\bibfnamefont {K.~S.}\ \bibnamefont
  {Turitsyn}}, \bibinfo {author} {\bibfnamefont {L.}~\bibnamefont {Lai}}, \
  and\ \bibinfo {author} {\bibfnamefont {W.~W.}\ \bibnamefont {Zhang}},\
  }\href@noop {} {\bibfield  {journal} {\bibinfo  {journal} {Phys. Rev. Lett}\
  }\textbf {\bibinfo {volume} {103}},\ \bibinfo {pages} {124501} (\bibinfo
  {year} {2009})}\BibitemShut {NoStop}%
\bibitem [{\citenamefont {Bridges}(2009)}]{bridges:09}%
  \BibitemOpen
  \bibfield  {author} {\bibinfo {author} {\bibfnamefont {T.~J.}\ \bibnamefont
  {Bridges}},\ }\href@noop {} {\bibfield  {journal} {\bibinfo  {journal}
  {Nonlinearity}\ }\textbf {\bibinfo {volume} {22}},\ \bibinfo {pages} {947}
  (\bibinfo {year} {2009})}\BibitemShut {NoStop}%
\bibitem [{\citenamefont {Gandzha}\ and\ \citenamefont
  {Lukomsky}(2007)}]{gandzha:07}%
  \BibitemOpen
  \bibfield  {author} {\bibinfo {author} {\bibfnamefont {I.~S.}\ \bibnamefont
  {Gandzha}}\ and\ \bibinfo {author} {\bibfnamefont {V.~P.}\ \bibnamefont
  {Lukomsky}},\ }\href@noop {} {\bibfield  {journal} {\bibinfo  {journal}
  {Proc. R. Soc. A}\ }\textbf {\bibinfo {volume} {463}},\ \bibinfo {pages}
  {1597} (\bibinfo {year} {2007})}\BibitemShut {NoStop}%
\bibitem [{\citenamefont {Amick}\ \emph {et~al.}(1982)\citenamefont {Amick},
  \citenamefont {Fraenkel},\ and\ \citenamefont {Toland}}]{amick:82}%
  \BibitemOpen
  \bibfield  {author} {\bibinfo {author} {\bibfnamefont {C.~J.}\ \bibnamefont
  {Amick}}, \bibinfo {author} {\bibfnamefont {L.~E.}\ \bibnamefont {Fraenkel}},
  \ and\ \bibinfo {author} {\bibfnamefont {J.~F.}\ \bibnamefont {Toland}},\
  }\href@noop {} {\bibfield  {journal} {\bibinfo  {journal} {Acta Math.}\
  }\textbf {\bibinfo {volume} {148}},\ \bibinfo {pages} {193} (\bibinfo {year}
  {1982})}\BibitemShut {NoStop}%
\bibitem [{\citenamefont {Longuet-Higgins}\ and\ \citenamefont
  {Fox}(1977)}]{lhf:77}%
  \BibitemOpen
  \bibfield  {author} {\bibinfo {author} {\bibfnamefont {M.~S.}\ \bibnamefont
  {Longuet-Higgins}}\ and\ \bibinfo {author} {\bibfnamefont {M.~J.~H.}\
  \bibnamefont {Fox}},\ }\href@noop {} {\bibfield  {journal} {\bibinfo
  {journal} {J. Fluid Mech.}\ }\textbf {\bibinfo {volume} {80}},\ \bibinfo
  {pages} {721} (\bibinfo {year} {1977})}\BibitemShut {NoStop}%
\bibitem [{\citenamefont {Longuet-Higgins}\ and\ \citenamefont
  {Fox}(1978)}]{lhf:78}%
  \BibitemOpen
  \bibfield  {author} {\bibinfo {author} {\bibfnamefont {M.~S.}\ \bibnamefont
  {Longuet-Higgins}}\ and\ \bibinfo {author} {\bibfnamefont {M.~J.~H.}\
  \bibnamefont {Fox}},\ }\href@noop {} {\bibfield  {journal} {\bibinfo
  {journal} {J. Fluid Mech.}\ }\textbf {\bibinfo {volume} {85}},\ \bibinfo
  {pages} {769} (\bibinfo {year} {1978})}\BibitemShut {NoStop}%
\bibitem [{\citenamefont {Penney}\ and\ \citenamefont
  {Price}(1952)}]{penney:52}%
  \BibitemOpen
  \bibfield  {author} {\bibinfo {author} {\bibfnamefont {W.~G.}\ \bibnamefont
  {Penney}}\ and\ \bibinfo {author} {\bibfnamefont {A.~T.}\ \bibnamefont
  {Price}},\ }\href@noop {} {\bibfield  {journal} {\bibinfo  {journal} {Phil.
  Trans. R. Soc. London A}\ }\textbf {\bibinfo {volume} {244}},\ \bibinfo
  {pages} {254} (\bibinfo {year} {1952})}\BibitemShut {NoStop}%
\bibitem [{\citenamefont {Iooss}\ \emph {et~al.}(2005)\citenamefont {Iooss},
  \citenamefont {Plotnikov},\ and\ \citenamefont {Toland}}]{iooss05}%
  \BibitemOpen
  \bibfield  {author} {\bibinfo {author} {\bibfnamefont {G.}~\bibnamefont
  {Iooss}}, \bibinfo {author} {\bibfnamefont {P.~I.}\ \bibnamefont
  {Plotnikov}}, \ and\ \bibinfo {author} {\bibfnamefont {J.~F.}\ \bibnamefont
  {Toland}},\ }\href@noop {} {\bibfield  {journal} {\bibinfo  {journal} {Arch.
  Rat. Mech. Anal.}\ }\textbf {\bibinfo {volume} {177}},\ \bibinfo {pages}
  {367} (\bibinfo {year} {2005})}\BibitemShut {NoStop}%
\bibitem [{\citenamefont {Mercer}\ and\ \citenamefont
  {Roberts}(1992)}]{mercer:92}%
  \BibitemOpen
  \bibfield  {author} {\bibinfo {author} {\bibfnamefont {G.~N.}\ \bibnamefont
  {Mercer}}\ and\ \bibinfo {author} {\bibfnamefont {A.~J.}\ \bibnamefont
  {Roberts}},\ }\href@noop {} {\bibfield  {journal} {\bibinfo  {journal} {Phys.
  Fluids A}\ }\textbf {\bibinfo {volume} {4}},\ \bibinfo {pages} {259}
  (\bibinfo {year} {1992})}\BibitemShut {NoStop}%
\bibitem [{\citenamefont {Wu}(1997)}]{wu:97}%
  \BibitemOpen
  \bibfield  {author} {\bibinfo {author} {\bibfnamefont {S.}~\bibnamefont
  {Wu}},\ }\href@noop {} {\bibfield  {journal} {\bibinfo  {journal} {Invent.
  Math.}\ }\textbf {\bibinfo {volume} {130}},\ \bibinfo {pages} {39} (\bibinfo
  {year} {1997})}\BibitemShut {NoStop}%
\bibitem [{\citenamefont {Taylor}(1953)}]{taylor:53}%
  \BibitemOpen
  \bibfield  {author} {\bibinfo {author} {\bibfnamefont {G.~I.}\ \bibnamefont
  {Taylor}},\ }\href@noop {} {\bibfield  {journal} {\bibinfo  {journal} {Proc.
  Roy. Soc. A}\ }\textbf {\bibinfo {volume} {218}},\ \bibinfo {pages} {44}
  (\bibinfo {year} {1953})}\BibitemShut {NoStop}%
\bibitem [{\citenamefont {Grant}(1973)}]{grant}%
  \BibitemOpen
  \bibfield  {author} {\bibinfo {author} {\bibfnamefont {M.~A.}\ \bibnamefont
  {Grant}},\ }\href@noop {} {\bibfield  {journal} {\bibinfo  {journal} {J.
  Fluid Mech.}\ }\textbf {\bibinfo {volume} {60}},\ \bibinfo {pages} {593}
  (\bibinfo {year} {1973})}\BibitemShut {NoStop}%
\bibitem [{\citenamefont {Okamura}(1998)}]{okamura:98}%
  \BibitemOpen
  \bibfield  {author} {\bibinfo {author} {\bibfnamefont {M.}~\bibnamefont
  {Okamura}},\ }\href@noop {} {\bibfield  {journal} {\bibinfo  {journal} {Wave
  Motion}\ }\textbf {\bibinfo {volume} {28}},\ \bibinfo {pages} {79} (\bibinfo
  {year} {1998})}\BibitemShut {NoStop}%
\bibitem [{\citenamefont {Okamura}(2003)}]{okamura:03}%
  \BibitemOpen
  \bibfield  {author} {\bibinfo {author} {\bibfnamefont {M.}~\bibnamefont
  {Okamura}},\ }\href@noop {} {\bibfield  {journal} {\bibinfo  {journal} {Wave
  Motion}\ }\textbf {\bibinfo {volume} {37}},\ \bibinfo {pages} {173} (\bibinfo
  {year} {2003})}\BibitemShut {NoStop}%
\bibitem [{\citenamefont {Okamura}(2010)}]{okamura:10}%
  \BibitemOpen
  \bibfield  {author} {\bibinfo {author} {\bibfnamefont {M.}~\bibnamefont
  {Okamura}},\ }\href@noop {} {\bibfield  {journal} {\bibinfo  {journal} {J.
  Fluid Mech.}\ }\textbf {\bibinfo {volume} {646}},\ \bibinfo {pages} {481}
  (\bibinfo {year} {2010})}\BibitemShut {NoStop}%
\bibitem [{\citenamefont {Schultz}\ \emph {et~al.}(1998)\citenamefont
  {Schultz}, \citenamefont {Vanden-Broeck}, \citenamefont {Jiang},\ and\
  \citenamefont {Perlin}}]{schultz}%
  \BibitemOpen
  \bibfield  {author} {\bibinfo {author} {\bibfnamefont {W.~W.}\ \bibnamefont
  {Schultz}}, \bibinfo {author} {\bibfnamefont {J.-M.}\ \bibnamefont
  {Vanden-Broeck}}, \bibinfo {author} {\bibfnamefont {L.}~\bibnamefont
  {Jiang}}, \ and\ \bibinfo {author} {\bibfnamefont {M.}~\bibnamefont
  {Perlin}},\ }\href@noop {} {\bibfield  {journal} {\bibinfo  {journal} {J.
  Fluid Mech.}\ }\textbf {\bibinfo {volume} {369}},\ \bibinfo {pages} {253}
  (\bibinfo {year} {1998})}\BibitemShut {NoStop}%
\bibitem [{\citenamefont {Wilkening}\ and\ \citenamefont {Yu}()}]{jia:11}%
  \BibitemOpen
  \bibfield  {author} {\bibinfo {author} {\bibfnamefont {J.}~\bibnamefont
  {Wilkening}}\ and\ \bibinfo {author} {\bibfnamefont {J.}~\bibnamefont {Yu}},\
  }\href@noop {} {\ }\bibinfo {note} {(in preparation)}\BibitemShut {NoStop}%
\bibitem [{\citenamefont {Ambrose}\ and\ \citenamefont
  {Wilkening}(2010)}]{vtxs1}%
  \BibitemOpen
  \bibfield  {author} {\bibinfo {author} {\bibfnamefont {D.~M.}\ \bibnamefont
  {Ambrose}}\ and\ \bibinfo {author} {\bibfnamefont {J.}~\bibnamefont
  {Wilkening}},\ }\href@noop {} {\bibfield  {journal} {\bibinfo  {journal}
  {Proc. Nat. Acad. Sci.}\ }\textbf {\bibinfo {volume} {107}},\ \bibinfo
  {pages} {3361} (\bibinfo {year} {2010})}\BibitemShut {NoStop}%
\bibitem [{\citenamefont {Nocedal}\ and\ \citenamefont
  {Wright}(1999)}]{nocedal}%
  \BibitemOpen
  \bibfield  {author} {\bibinfo {author} {\bibfnamefont {J.}~\bibnamefont
  {Nocedal}}\ and\ \bibinfo {author} {\bibfnamefont {S.~J.}\ \bibnamefont
  {Wright}},\ }\href@noop {} {\emph {\bibinfo {title} {Numerical
  Optimization}}}\ (\bibinfo  {publisher} {Springer},\ \bibinfo {address} {New
  York},\ \bibinfo {year} {1999})\BibitemShut {NoStop}%
\bibitem [{\citenamefont {Keller}(1987)}]{keller:87}%
  \BibitemOpen
  \bibfield  {author} {\bibinfo {author} {\bibfnamefont {H.~B.}\ \bibnamefont
  {Keller}},\ }\href@noop {} {\emph {\bibinfo {title} {Numerical methods in
  bifurcation problems}}}\ (\bibinfo  {publisher} {Springer},\ \bibinfo
  {address} {New York},\ \bibinfo {year} {1987})\BibitemShut {NoStop}%
\bibitem [{\citenamefont {Mercer}\ and\ \citenamefont
  {Roberts}(1994)}]{mercer:94}%
  \BibitemOpen
  \bibfield  {author} {\bibinfo {author} {\bibfnamefont {G.~N.}\ \bibnamefont
  {Mercer}}\ and\ \bibinfo {author} {\bibfnamefont {A.~J.}\ \bibnamefont
  {Roberts}},\ }\href@noop {} {\bibfield  {journal} {\bibinfo  {journal} {Wave
  Motion}\ }\textbf {\bibinfo {volume} {19}},\ \bibinfo {pages} {233} (\bibinfo
  {year} {1994})}\BibitemShut {NoStop}%
\bibitem [{\citenamefont {Smith}\ and\ \citenamefont
  {Roberts}(1999)}]{smith:roberts:99}%
  \BibitemOpen
  \bibfield  {author} {\bibinfo {author} {\bibfnamefont {D.~H.}\ \bibnamefont
  {Smith}}\ and\ \bibinfo {author} {\bibfnamefont {A.~J.}\ \bibnamefont
  {Roberts}},\ }\href@noop {} {\bibfield  {journal} {\bibinfo  {journal} {Phys.
  Fluids}\ }\textbf {\bibinfo {volume} {11}},\ \bibinfo {pages} {1051}
  (\bibinfo {year} {1999})}\BibitemShut {NoStop}%
\end{thebibliography}
%

\end{document}